\documentclass[11pt]{article}
\usepackage{graphicx}
\usepackage{epstopdf} %needs to be loaded after graphicx
\usepackage{helvet}
\usepackage{setspace}
\usepackage{rotating}
\usepackage{booktabs}
\usepackage{lscape}
\usepackage[superscript]{cite}
\usepackage{caption}
\usepackage{longtable}
\usepackage{color}
\begin{document}
\thispagestyle{empty}
\fontfamily{phv}\selectfont{%Helvetica
%{\huge \noindent x}
{\LARGE \noindent A new interpretation of quantum theory, based on a bundle-theoretic view of objective idealism}
\\
\\
{\scriptsize Martin Korth$^{*}$\\
%$^{1}$
IVV NWZ, WWU Münster, Wilhelm-Klemm-Str. 10, 48149 Münster (Germany)\\
$^{*}$ Corresponding author email: dgd@uni-muenster.de\\
v1, 2022/08/22}\\ \\

%\noindent {\bf \fontfamily{phv}\selectfont{Keywords:}} \\ \\

%\noindent {\bf \fontfamily{phv}\selectfont{Abstract}}\\
%\noindent x
%\\ \\

%\noindent {\bf \fontfamily{phv}\selectfont{Introduction}}\\
\noindent
After about a century since the first attempts by Bohr, the interpretation of quantum theory is still a field with many open questions.\cite{Zeilinger}
%A snapshot of foundational attitudes toward quantum mechanics
%MaximilianSchlosshauer, Johannes Kofler, AntonZeilinger
%Studies in History and Philosophy of Science Part B: Studies in History and Philosophy of Modern Physics
%Volume 44, Issue 3, August 2013, Pages 222-230, 
In this article a new interpretation of quantum theory is suggested, motivated by philosophical considerations. Based on the findings that the 'weirdness' of quantum theory can be understood to derive from a vanishing distinguishability of indiscernible particles, and the observation that a similar vanishing distinguishability is found for bundle theories in philosophical ontology, the claim is made that quantum theory can be interpreted in an intelligible way by positing a bundle-theoretic view of objective idealism instead of materialism as the underlying fundamental nature of reality.
\\ \\

\noindent {\bf \fontfamily{phv}\selectfont{Open questions in the interpretation of quantum theory}}\\
\noindent 
In quantum theory, upon the formation of a system from parts, e.g. a molecule from elementary particles, the constituting parts are taken up into the whole system in a way which makes them loose their independent identity (discussed in Philosophy as `quantum non-individuality'), so that the system requires some global description, e.g. via a wave function and its evolution, until measurement occurs, which irreversibly collapses this construction to give statistically distributed and quantized (i.e. not continuously distributed) results.

Although the correctness and usefulness of quantum theory is beyond any doubt, most scientist seem to agree that it is still quite unclear how to make sense of it within our common materialist scientific world view. The open questions can be grouped into four fractions, the first of which considers the basic observations of quantized (particle-like) interactions, indeterminism (the statistical nature of results), and the possibility of transmutation (the conversion of elementary particles into each other).\cite{Isham}
%C. J. Isham. Lectures on Quantum Theory (London: Imperial College Press, 2001).
The second fraction is about questions related to the 'holistic' nature of quantum systems: Why is there no identity of indiscernibles (shown e.g. by quantum statistics), but complementarity and uncertainty (i.e. that certain properties cannot be measured simultaneously)? Why non-locality (shown by Bell test experiments)\cite{Bell}
%J. S. Bell, On the problem of hidden variables in quantum mechanics. Rev. Mod. Phys. 38, 447–452 (1966).
 and entanglement (non-local coupling of properties)? The third fraction deals with the probably most central question: Why does the act of measurement have a special role in an objective scientific theory? And why is measurement not just a revelation of pre-existing values, but depends on the context of measurement?\cite{KochenSpecher}
%S. Kochen, E. P. Specker, The problem of hidden variables in quantum mechanics. J. Math. Mech. 17, 59–87 (1967).
Within a forth section one could collect all the more technical questions: Why does the mathematical machinery need complex numbers? Wherefrom the parameters of the standard model etc.

I will argue in the following that the vanishing distinguishability of indiscernible particles within micro-scale systems is at the heart of quantum theory: Consider a system in which the properties of indiscernible contributing parts are really taken up by the system until parts are forced into distinction, upon which the system looses those properties, which need to be assigned to allow for the formation of a proper distinct part. Such a system would clearly show non-locality and could show entanglement due to the system-wide nature of the accounting of properties. Depending on how properties are taken up and given away again by the system it could also show complementarity and uncertainty: It might not be possible to transfer certain properties at exactly the same time. The measurement act could be understood as one, indeed context-dependent way to force micro-scale distinction, but forced distinction would not be bound to a human observer,
thereby removing any need for a special role of the later. Even the mentioned more basic observations of quantization (properties come and go in batches), indeterminism (properties are lost, i.e. become manifest in former parts statistically) and transmutation (only the overall accounting of properties is tracked) can be understood this way. And this ansatz can also be discussed from the viewpoint of Feynman's path integral formulation of quantum theory: Because particle identity is indeed taken up by the super-system, we have to take into account infinitely many possible trajectories for them. (The necessary weighing of trajectories would then bring complex numbers into the mathematical machinery.)
%idealism allow for quantum computing via contextuality
%why complex? non-commutation/amplitude+phase->interference
%~ why born rule: double slit; not real prob. but complex prob. amplitude added
Concerning interpretations of quantum theory, the view put forward here has overlap with both ensemble interpretations and quantum decoherence.
In Philosophy, Morganti has argued similarly, that in quantum theory we should consider properties to belong to the whole system.\cite{Morgani09}
%TODO read more
%Inherent properties and statistics with individual particles in quantum mechanics
%MatteoMorganti
%Studies in History and Philosophy of Science Part B: Studies in History and Philosophy of Modern Physics
%Volume 40, Issue 3, August 2009, Pages 223-231
%TODO double slit experiment example?
%TODO meso-scale Gibb's mixing paradox also solvable with indistinguishability?

The proposed taking up of parts into a whole of course implies their -- at least temporary -- vanishing from the material world, which would constitute a major breach of etiquette within our common scientific world-view, in which the substance of objects is supposed to be of material nature (however fleeting our concept of matter has become with quantum field theory). It is after all the materialist background of our modern world view which made quantum theory look "weird" to us in the first place.
\\ \\

\noindent {\bf \fontfamily{phv}\selectfont{A bundle-theoretic view of objective idealism}}\\
\noindent
%NOPE
%A World of Universals, John O'Leary-Hawthorne and J. A. Cover, Philosophical Studies: An International Journal for Philosophy in the Analytic Tradition, Vol. 91, No. 3 (Sep., 1998), pp. 205-219 (15 pages)
Interestingly, a similar vanishing distinguishability of indiscernibles is discussed for the 'bundle theory' of objecthood in philosophical ontology, a theory which can hardly be re-framed against a materialist background, but lends itself to the development of objective idealist theories of reality. (Such theories are based on the assumption of an objective existence of non-material entities, as opposed to subjective idealism, in which the world is a mere consequence of subjective mental phenomena). Unlike in the rivaling 'substance theory', where objects are discussed as to be constituted by a substance which bornes properties, bundle theories assume objects to be no more than the bundle of its properties, without any so-called `bare particular' at its core to identify its essence under change. A major issue with bundle theories is that the so-called `compresence' relation, which constitutes the bundling of qualities, leads to a range of logical puzzles, with at least one puzzle hard to ignore when dealing with the material world: It can be shown that compresence can not account for a proper individuation of indiscernible objects, 
because objects with exactly the same bundle of universal qualities become essentially the same object.\cite{HS}
%Locations, John Hawthorne & Theodore Sider, Philosophical Topics 30 (1):53-76 (2002) 
And if position in space should be invoked to account for this problem, it is unclear how the compresence relation can do this without infinite regress, as some form of linkage seem to be required to make the spatial relations of indiscernible objects consistent (see Hawthorne and Sider\cite{HS}
%Locations, John Hawthorne & Theodore Sider, Philosophical Topics 30 (1):53-76 (2002) 
for further discussion). While usually considered a knockout argument against bundle theories, I will argue that vanishing distinguishability is not a bug but a core feature, allowing for an intelligible interpretation of quantum theory in the light of a scientifically tenable version of objective idealism.

I have elsewhere given an outline of how such an objective idealist world-view would have to look like\cite{MKscience},
%M. Korth, Towards a scientifically tenable description of objective idealism, forthcoming, PDF on http://mkorth.de
but the upshot is that modern science would have to be re-interpreted as the limiting case of a both material and non-material world. (Requiring no changes to current science at the material limit.) Objects would be bundles of non-material, but subject-independently existing building blocks of qualitative nature. At the very bottom, only spatial relations and elementary particle properties would make up the bundles, thereby giving rise to the material world, but at a higher level non-material properties might be part of bundles.
%GR Some variation of the model is the currently best formulation of idealism to engage with modern science
Indiscernible sub-bundles (like elementary particles) would become  indistinguishable within larger bundles (like a molecular system), thereby 'handing over' their properties to the overall system, quite as we have speculated for a physical system that needs to be described by a wave function. 
%TODO distinction through spatial relation; only direct micro-scale relations count?
Only that in the objective idealist world-view, the temporary vanishing of micro-scale entities would not be problematic, as the bundle keeps the constituting properties with no necessity for being itself constituted by separate sub-bundles. A wave function would then have to be understood as taking stock of the materially relevant properties of such a bundle.
\\ \\

%TODO read more
\noindent {\bf \fontfamily{phv}\selectfont{Quantum Reconstruction}}\\
\noindent
The last two decades have seen an increasing interest in the project of `quantum reconstruction',
i.e. of deriving Quantum Theory from as simple as possible assumptions,
so that it is possible for us to compare our assumptions with the findings in this field.
Hardy's work awakened the interest, when he argued that the core trait of quantum theory is its inherently probabilistic nature,
and then showed that the simplest possible theory for this is quantum mechanics.\cite{Hardy}
%Quantum Theory From Five Reasonable Axioms, Lucien Hardy,  	arXiv:quant-ph/0101012, 2001
%TODO adding up of propability amplitudes as probabilities, not densities
% Schrödinger: amplitude = smeared-out density vs Born: amplitude = probability
% -> Particle is not smeared out, but taken up; OK!
More recently, Masanes, Galley, and Müller have found that starting from the set of assumptions
that we have to consider the case of measuring unique values from unitarily (smoothly) evolving quantum states,
we automatically arrive at the Born rule, which connects the mathematical mechanism of Quantum Theory
with the interpretations of the outcomes.\cite{Masanes} %PLUS: partioning plays no role!
%Masanes, L., Galley, T.D. & Müller, M.P. The measurement postulates of quantum mechanics are operationally redundant. Nat Commun 10, 1361 (2019). https://doi.org/10.1038/s41467-019-09348-x
%HERE ?
%Hidden in their set of assumptions is the property of quantum states to obey quantum statistics;
%a mathematical formulation of the indistinguishability of indiscernble particles.
Cabello proposed that there is no underlying physical law for measurement outcomes,
but only a set of consistency requirements which have to be met;
the Born rule is then just the outcome of these requirements.\cite{Cabello}
%Quantum correlations from simple assumptions, Adán Cabello, Phys. Rev. A 100, 032120 – Published 24 September 2019
Recent work in quantum reconstruction thus seems to be very much in line with our new model,
insofar as it tells us that quantum theory is of inherent probabilistic nature,
and that the Born rule derives naturally from this,
with only certain overall consistency requirements to be fulfilled
(i.e. the overall accounting of properties in the bundle).
\\ \\

\noindent {\bf \fontfamily{phv}\selectfont{Conclusions}}\\
\noindent The above outlined argument that a bundle-theoretic view of objective idealism might allow for a comprehensible interpretation of quantum theory
gives in my opinion first of all further support to the project of investigating the possibility of scientifically tenable idealistic theories.

This should not distract from the fact that the above proposal for a new interpretation of quantum theory itself surely needs further work. % to explain for instance mixed-particle systems.
%TODO can they become indistinguishable at the level of their respective super-systems?
Concerning for instance one particle systems, one would have to assume that the particle
interacts with an earlier versions of itself. This is well in line with the bundle-theoretic view,
where matter moves by the transfer of particle properties from one space-point to the next.
If properties are not first removed and then added, but first added and the removed,
each particle will indeed always see (at least) one past instance of itself.
%TODO exchange?
(I will further explore movement and relativity in objective idealism within a separate article.)
%GR Lorenz-Gleilchungen+Deutung je mehr spacepoints verschmiert

Interestingly, the above given interpretation of quantum theory is not a one-sided project,
as in return the findings of quantum theory give support to bundle over substance theories
within the philosophical discussion of objecthood.
Finally, the most important `prediction' of the new interpretation would be that 
wave functions collapse whenever indiscernible parts become distinguishable.
\\ \\

%\noindent {\bf \fontfamily{phv}\selectfont{Acknowledgments}}\\
%\noindent The author would like to thank 
%HERE
%\textcolor{red}{ %ALERT
%This work was supported in part by 
%}
%\\

%\noindent {\bf \fontfamily{phv}\selectfont{Author ontributions}}\\ \\
%\noindent Both authors contributed equally to all parts of the work.
%\\

%\noindent {\bf \fontfamily{phv}\selectfont{Supplementary Information}}\\
%\noindent Supplementary Information is available online at ...
%Raw data will be made available as a web-accessible database on http://qmcathome.org/clean\_mobility\_now.html.
%\\

%\bibliographystyle{achemsolx2}
%\fontfamily{phv}\selectfont{\bibliography{bib_paper}}

}%Helvetica
\end{document}